\documentclass[]{pasj01}

\Received{}
\Accepted{}

\begin{document}

\title{
Periodically-repeating fast radio bursts: Lense-Thirring precession of a debris disk?}

\author{Wen-Cong \textsc{Chen}}
\altaffiltext{}{School of Science, Qingdao University of Technology, Qingdao 266525, China;\\
School of Physics and Electrical Information, Shangqiu Normal University, Shangqiu 476000, China}
\email{chenwc@pku.edu.cn}


\maketitle

\begin{abstract}
Recently, repeating fast radio bursts (FRBs) with a period of $P_{\rm FRB}=16.35\pm0.18$ days from FRB 180916.J0158+65 had been reported. It still remains controversial how to give rise to such a periodicity of this FRB. In this Letter, based on an assumption of a young pulsar surrounding by a debris disk, we attempt to diagnose whether the Lense-Thirring precession of the disk on the emitter can produce the observed periodicity. Our calculations indicate that the Lense-Thirring effect of a tilted disk can result in a precession period of 16 days for a mass inflow rate of $0.5-1.5\times10^{18}~\rm g\,s^{-1}$, a spin period of 1-20 ms of the pulsar, and an extremely low viscous parameter $\alpha=10^{-8}$ in the disk. The disk mass and the magnetic field of the pulsar are also constrained to be $\sim10^{-3}~\rm M_{\odot}$ and $< 2.5\times 10^{13}~\rm G$. In our model, a new born pulsar with normal magnetic field and millisecond period would successively experience accretion phase, propeller phase, and is visible as a strong radio source in the current stage. The rotational energy of such a young NS can provide the observed radio bursting luminosity for $400$ years.
\end{abstract}

\section{Introduction}

Fast radio bursts (FRBs) are intriguing  GHz radio transients with typical durations of a handful of milliseconds \citep{lori07,thor13,burk14,spit14,masu15,ravi16}. According to their anomalously high dispersion measures (DM $\sim 300-1500~\rm pc\,cm^{-3}$), FRBs were widely thought to be bursting events at cosmological distance. Especially, the detection of the first repeating FRB 121102 proved its cosmological origin \citep{chat17,marc17,tend17}.

At present, the physical origin of FRBs still remains controversial. Recently, the discovery of periodic activity from FRB 180916.J0158+65 raised a lively interest of many astrophysicists in understanding the physical nature of repeating FRBs \citep{chim20}. From September 2018 to November 2019, this source was observed twenty eight bursts, which emerges a periodicity of
$P_{\rm FRB}=16.35\pm0.18$ days.

What is the producing mechanism of such a periodicity? The first probability is the orbital motion of a binary. \citet{ioka20} proposed an interesting binary comb model, in which the strong wind of the companion exerts a comb on the FRB
pulsar \citep{zhang17,zhang18}, and the comb can modulate emission beams of the FRB to synchronize with the orbital motion of the binary system. \citet{lyut20} discussed a similar model, the periodicity of the FRB originates from free-free absorption in the wind of an OB primary that orbits a NS. Other binary models with different bursting mechanisms such as the encounters between the pulsar and asteroids \citep{dai16,dai20}, the interaction of smaller bodies with a pulsar wind \citep{mott20}, accretion of a neutron star (NS) from a magnetized white dwarf \citep{gu20} were also proposed. The second mechanism producing periodicity is the precession of the emitter. \citet{yang20} argued that the periodicity of the FRB arises from a spin precession of the emitting NS induced by the orbital motion. Furthermore, the free precession of magnetar \citep{levi20,zana20} and the forced precession of an isolated magnetar by a fallback disk \citep{tong20} were also employed to account for the periodicity of the FRB.

Actually, Lense-Thirring precession can also result in a periodic variation. \citet{sara80} found that Lense-Thirring precession of the compact object in SS 433 by a moderately massive disk can interpret 164 days period of Doppler-shifted lines from H I and He I. Based on the relativistic dragging of inertial frames surrounding an accreting NS, \citet{stel98} showed that Lense-Thirring precession can produce kHz quasi-periodic oscillation in the X-ray light curves of low-mass X-ray binaries. Recently, Lense-Thirring precession of the orbit originating from rapid rotation of a white dwarf was confirmed in the binary pulsar PSR J1141-6545 \citep{kris20}.

Assuming isolated NS with millisecond period as the origin of the FRB, in this Letter we intend to diagnose whether a Lense-Thirring precession of the NS caused by a fallback disk can be responsible for the periodicity of $P=16.35\pm0.18$ days. We describe the Lense-Thirring procession model of a debris disk in section 2, and present the summary and discussion in section 3.

\section{Disk Lense-Thirring precession model}
Lense-Thirring precession comes from the rotation of inertial frames in the proximity of a rotating object. It will cause all conserved vectors including angular momentum in the corresponding zone precess at a very nearly same rotation rate.  A misaligned accretion disk surrounding rotating central NS or black hole was expected to be exerted on a torque by the Lense-Thirring effect, which causes the precession of the inner disk \citep{bard75}. On the contrary, if the part of disk possess an angular momentum more than the central compact object, it will precess the compact object \citep{sara80}. In this work, we propose that the tilted fallback disk around a young NS with millisecond period causes the precession of radio bursting beam, and gives rise to a periodicity of FRBs.

\subsection{precession radius}
In the disk, the mass inflow rate at the radius $r$ is $\dot{M}=2\pi r v_{\rm r}\Sigma$, where $\Sigma$ is the surface density of the disk and $v_{r}$ is the radial inflow velocity. The angular momentum of a mini ring with a differential width $dr$ in the disk is $dJ=2\pi r^{2}\Sigma v_{\varphi}dr$ \citep{lu05}, where $v_{\varphi}$ is the rotation velocity of the disk. Therefore,
the angular momentum per logarithmic interval of the radius is given by \citep{sara80}
\begin{equation}
J(r)=\frac{dJ}{d({\rm ln}r)}=\dot{M}r^{2}\left(\frac{v_{\varphi}}{v_{r}}\right).
\end{equation}

Assuming that the total angular momentum of the disk is greater than that of the NS $J_{\rm ns}$, there exist a critical radius at which $J(r)=J_{\rm ns}$, i. e. the precession radius $r_{\rm p}$ \citep{sara80}. The outside disk with $r>r_{\rm p}$ will precess the central NS and the inner disk with $r<r_{\rm p}$ through the gravitational coupling between the central object and the disk. The angular momentum of the NS is $J_{\rm ns}=2\pi I/P$, here $I$ is the moment of inertia (we take $I=10^{45}~\rm g\,cm^{2}$), $P$ is the spin period. The precession radius can be written as
\begin{equation}
r_{\rm p}=\sqrt{\frac{J_{\rm ns}v_{\rm r}}{\dot{M}v_{\varphi}}}\approx7.9\times 10^{13}\dot{M}_{18}^{-1/2}P^{-1/2}\left(\frac{v_{r}}{v_{\varphi}}\right)^{1/2}~~\rm cm,
\end{equation}
where $\dot{M}=10^{18}~\rm g\,s^{-1}\dot{M}_{18}$, $P$ in units of second.

Assuming a standard thin disk, and taking the mass and the radius of NS to be $M=1.4~M_{\odot}$ and $R=10$ km, the radial inflow velocity at $r_{\rm p}$ is \citep{shak73}
\begin{equation}
v_{\rm r}\approx 1.96\times 10^{10}\alpha\dot{M}_{18}^{2}\left(\frac{r_{\rm p}}{10^{6}~\rm cm}\right)^{-5/2}~~\rm cm\,s^{-1},
\end{equation}
where $\alpha$ is the viscous parameter \footnote{We adopt a solution of the zone that is dominant by the radiation pressure, and take $1-(\frac{r}{3R_{\rm g}})^{-1/2}\approx1$ because $r\gg R_{\rm g}$, where $R_{\rm g}$ is the gravitational radius of the NS \citep{shak73}.}.
The disk is thought to obey the Keplerian motion, hence the rotation velocity at $r_{\rm p}$ is
\begin{equation}
v_{\varphi}=\sqrt{\frac{GM}{r_{\rm p}}}\approx 1.4\times 10^{10}\left(\frac{r_{\rm p}}{10^{6}~\rm cm}\right)^{-1/2}~~\rm cm\,s^{-1},
\end{equation}
where $G$ is the gravitational constant. According to equations (2), (3), and (4), we can get
\begin{equation}
\frac{v_{\rm r}}{v_{\varphi}}\approx 1.5\times 10^{-8}\alpha^{1/2}\dot{M}_{18}^{3/2}P^{1/2}.
\end{equation}
Inserting equation (5) into equation (2), the precession radius only depends on the mass inflow rate, the spin period of the NS, and the viscous parameter, hence it can be expressed as
\begin{equation}
r_{\rm p}=3.6\times 10^{8}\alpha_{-8}^{1/4}\dot{M}_{18}^{1/4}P_{\rm 5ms}^{-1/4}~~\rm cm,
\end{equation}
where $\alpha=10^{-8}\alpha_{-8}$, and $P=5~{\rm ms}~P_{\rm 5ms}$.

\subsection{precession period}

Because the fastest precession rate arises from the ring with $r=r_{\rm p}$, the Lense-Thirring effect can result in a slow precession of the NS and the inner disk at a precession period \citep{wilk72}
\begin{equation}
P_{\rm LT}=\frac{\pi c^{2}r_{\rm p}^{3}}{GJ_{\rm ns}}.
\end{equation}
Numerically, rewrite the precession period as
\begin{equation}
P_{\rm LT}=18.2\alpha^{3/4}_{-8}\dot{M}_{18}^{3/4}P^{1/4}_{5\rm ms}~~\rm days.
\end{equation}
Figure 1 illustrates the relation between the precession period and the mass inflow rate at the precession radius (we take a constant viscous parameter $\alpha=10^{-8}$). The parameter space producing the periodicity detected from FRB 180916.J0158+65 by the Lense-Thirring precess of a disk is very narrow. For the spin period $1-20$ ms of the NS, the mass inflow rate in the disk should be in the range of $0.5-1.5\times 10^{18}~\rm g\,s^{-1}$. Comparing with the mass inflow rate, the precession period is insensitive to the spin period because $P_{\rm LT}\propto P^{1/4}$. A large spin period tend to accompany by a low mass inflow rate. Similar to \cite{sara80}, our model requires an extremely low viscous parameter $\alpha=10^{-8}$.

To sustain a stable precession of the NS and the inner disk, the precession time ($P_{\rm LT}$) should be less than the viscous timescale at the precession radius \citep{kuma85,sche96,nata99}. The viscous timescale at $r_{\rm p}$ is \citep{armi99}
\begin{equation}
t_{\rm vi}=\frac{r_{\rm p}^{2}}{\alpha\Omega H_{\rm p}^{2}},
\end{equation}
where $\Omega$ and $H_{\rm p}$ are the angular velocity and the scale height of the disk at $r_{\rm p}$. At the precession radius $\Omega=v_{\varphi}/r_{\rm p}$, so $t_{\rm vi}=r_{\rm p}^{3}/\alpha v_{\varphi} H_{\rm p}^{2}$. We estimate the scale height by the thickness of the disk \citep{shak73}, so
\begin{equation}
\frac{H_{\rm p}}{r_{\rm p}}\sim 0.01\alpha_{-8}^{-1/4}\dot{M}_{18}^{3/4}P_{5\rm ms}^{1/4}.
\end{equation}
For some typical parameters, our estimated ratio between the scale height and the precession radius is about 0.01, which is sightly smaller than the value $H_{\rm p}/r_{\rm p}\approx0.03$ in the hydrostatically supported and geometrically thin disk  \citep{bell04}, Numerically, the viscous timescale is given by
\begin{equation}
t_{\rm vi}\approx 1.5\times 10^{4} \alpha_{-8}^{-5/8}\dot{M}_{18}^{3/8}P_{5\rm ms}^{-3/8}\left(\frac{H_{\rm p}/r_{\rm p}}{0.01}\right)^{-2}~~\rm years.
\end{equation}
For some typical parameters, the viscous timescale can be expected to be much greater than the precession time.

\subsection{debris disk mass}
In the fallback disk, the mass per logarithmic radius interval $M(r)=r\dot{M}/v_{\rm r}$ \citep{sara80}. We can estimate the disk mass as follows
\begin{equation}
M(r_{\rm p})=\frac{\dot{M}r_{\rm p}}{v_{\rm r}}\approx 2.3\times 10^{-3}\alpha_{-8}^{-1/8}\dot{M}_{18}^{-9/8}P_{\rm 5ms}^{-7/8}~~\rm M_{\odot}.
\end{equation}
To yield such a precession period of 16.35 days, a relatively massive disk ($ \sim 10^{-3}~M_{\odot}$) is expected.

According to the work about supernova fallback performed by \citet{cann90} and \citet{mine97}, the mass inflow rate in the disk can be roughly estimated to be
\begin{equation}
\dot{M}=\dot{M}_{0}\left(\frac{t_{\rm disk}}{1~\rm years}\right)^{-1.35}~{\rm g\,s^{-1}},
\end{equation}
where $t_{\rm disk}$ is the lifetime of the disk, $\dot{M}_{0}$ is the mass inflow rate when $t_{\rm disk}=1~\rm year$. In our model, a young fallback disk with $t_{\rm disk}\sim10^{3}~\rm years$ is required. According the mass inflow rate $\dot{M}_{18}\approx 1$ at the current stage, equation (13) yields $\dot{M}_{0}\approx 10^{22}~\rm g\,s^{-1}$.

\begin{figure}
\centering
\includegraphics[width=1.15\linewidth,trim={0 0 0 0},clip]{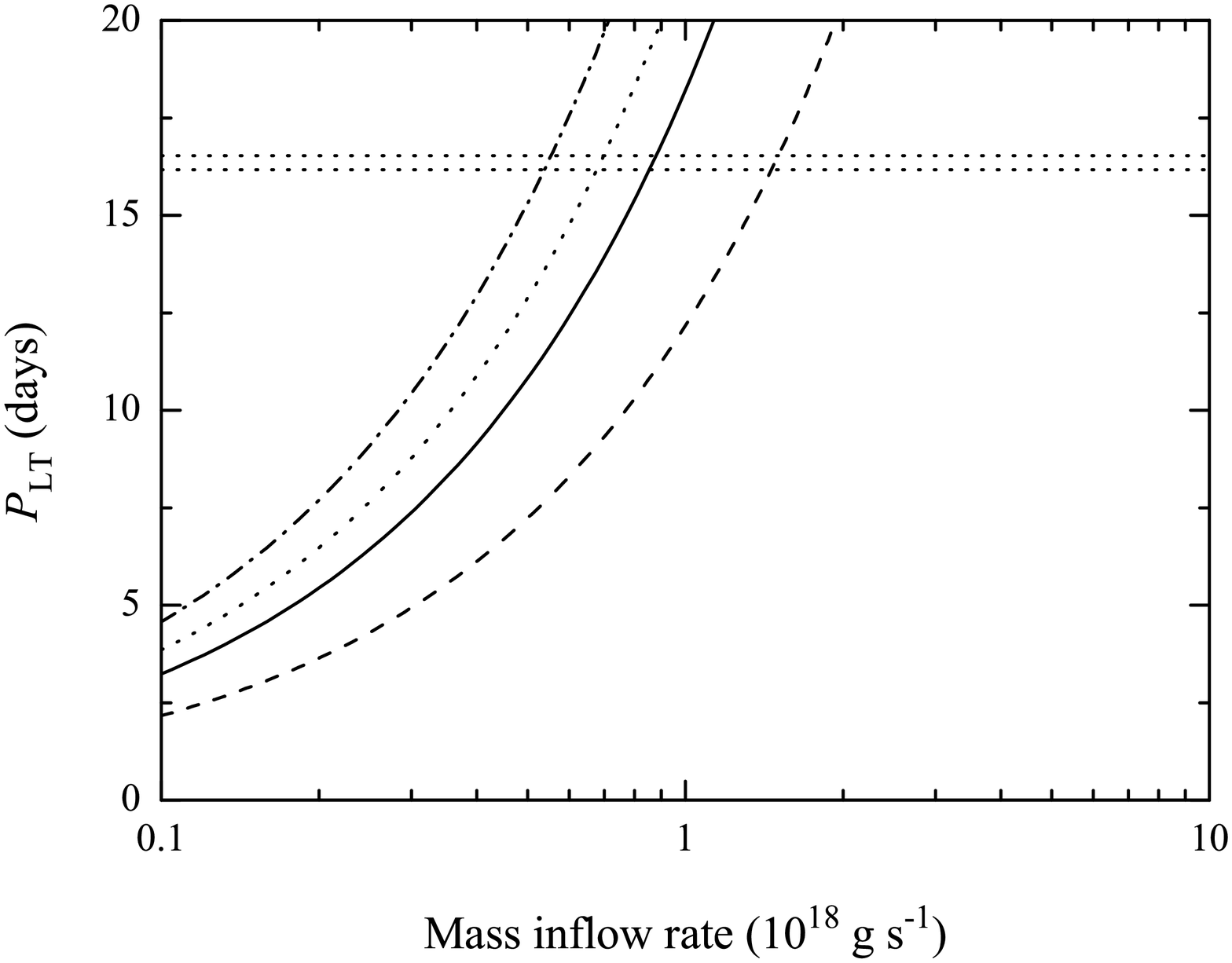}
\caption{Relation between the precession period and the mass inflow rate at the
precession radius. The solid, dashed, dotted, and dashed-dotted curves represent the spin period of the NS $P=5$, 1, 10, and 20 ms, respectively. Two horizontal dotted lines correspond to the detected periodicity of $16.35\pm0.18$ from FRB 180916.J0158+65.} \label{fig:orbmass}
\end{figure}

\subsection{Three radii}
It is customary to take the inner radius of the disk at the magnetospheric radius, which is defined by the equilibrium between the ram pressure of the infalling material and the magnetic pressure. The magnetospheric radius is \citep{davi73}
\begin{equation}
r_{\rm m}=\xi\left(\frac{\mu^{4}}{2GM\dot{M}^{2}}\right)^{1/7}\approx 8.3\times 10^{7}\dot{M}_{18}^{-2/7}\mu_{30}^{4/7}~\rm cm,
\end{equation}
where $\mu=BR^{3}/2$ ($B$ is the surface dipole magnetic field of the NS) the dipolar magnetic momentum of the NS, and we take the  dimensionless parameter $\xi=0.52$ \citep{ghos79}.

The precession of the central emitter by the outer disk requires $r_{\rm m}\leq r_{\rm p}$, so the surface magnetic field can be constrained to be
\begin{equation}
B \leq 2.6\times 10^{13}\alpha_{-8}^{7/16}\dot{M}_{18}^{15/16}P_{\rm 5ms}^{-7/16}~\rm G.
\end{equation}
In our model, the emitter of FRBs is a NS with normal magnetic field and millisecond period.

The corotation radius of the NS
\begin{equation}
r_{\rm co} =\sqrt[3]{\frac{GMP^{2}}{4\pi^{2}}}=4.9\times 10^{6}P_{\rm 5ms}^{2/3}~\rm cm,
\end{equation}
and the light cylinder radius
\begin{equation}
r_{\rm lc} =\frac{cP}{2\pi}=2.4\times 10^{7}P_{\rm 5ms}~\rm cm.
\end{equation}
Taking $P_{\rm 5ms}=1$, and $B=1.0\times 10^{12}~\rm G$, we plot the evolution of three radii in Figure 2. It is worth note that the corotation radius and the light cylinder radius should change with the age of the MSP. We only illustrate the transitions of three phases so that we ignore the variation of the spin period. When $t\sim 0-11$ years, a high mass inflow rate in the fallback disk results in $r_{\rm m}<r_{\rm co}$, a new born NS is in the accretion phase. Once $r_{\rm co}<r_{\rm m}<r_{\rm lc}$ ($t\sim11-110~ \rm years$), the NS enters the propeller phase. When $t> 110~\rm years$, $r_{\rm m}>r_{\rm lc}>r_{\rm co}$, therefore the NS at the current stage appears as radio pulsars, emitting radio radiation for $\sim 10^{3}~\rm years$ till the current stage.

\begin{figure}
\centering
\includegraphics[width=1.15\linewidth,trim={0 0 0 0},clip]{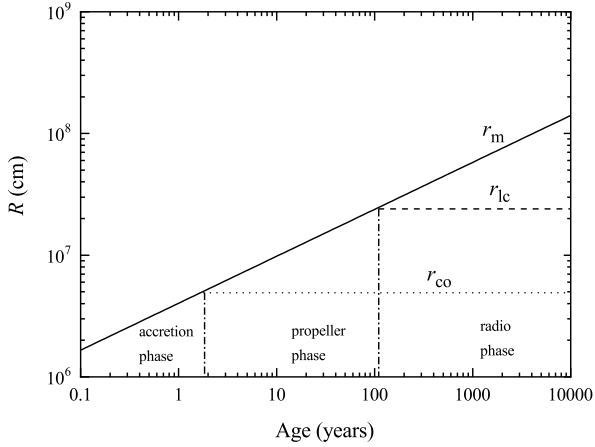}
\caption{Evolution of three radii of the pulsar. The solid, dashed, and dotted lines represent the magnetospheric radius, the light cylinder radius, and the corotation radius, respectively. Two vertical dashed-dotted lines correspond to the boundaries of three zones.} \label{fig:orbmass}
\end{figure}

\subsection{Energetics and rates}
In our model, the energy reservior of FRBs is the rotational energy of the NS. The rotational energy of a NS is $E_{\rm rot}=2\pi^{2}I/P^{2}\approx 7.9\times 10^{50}P_{\rm 5ms}^{-2}~\rm erg$, and the spin-down energy loss rate via magnetic dipole radiation is
\begin{eqnarray}
\dot{E}_{\rm sd} =\frac{32\pi^{4}B^{2}R^{6}}{3c^{3}P^{4}}\approx 6.2\times 10^{40}B_{12}^{2}P_{\rm 5ms}^{-4}{~\rm erg\,s^{-1}},
\end{eqnarray}
where $B=B_{12}10^{12}~\rm G$. Considering only a part ($\epsilon_{\rm r}$) of the rotational energy is emitted in radio waveband, and the radio luminosity of FRB 180916.J0158+65 in a single burst is $L_{\rm FRB}=3\times10^{40}~\rm erg\,s^{-1}$ \citep{chim20}, hence we have
\begin{equation}
L_{\rm FRB}=\dot{E}_{\rm sd}\epsilon_{\rm r}/f_{\rm b},
\end{equation}
where $f_{\rm b}$ is the beaming factor of the NS. Taking $f_{\rm b}=0.1$ \citep{muno20}, equation (19) yields a radio efficiency $\epsilon_{\rm r}\approx 0.05$,and the spin-down timescale
\begin{equation}
t_{\rm sd} =400 P_{\rm 5ms}^{2}B_{12}^{-2}~\rm years.
\end{equation}
For a young NS with a spin period of 5 ms, the spin-down energy can power the radio bursting luminosity of  FRB 180916.J0158+65 for 400 years.

We then estimate the rate of the FRB detections. The rate density of core collapse supernova (CCSN, the progenitors of NSs) in the local universe $R_{\rm CCSN}\approx 10^{5}~\rm Gpc^{-3}year^{-1}$ \citep{tayl14}, and the fraction of CCSN that result in the young highly spinning NSs $f = 0.1$ \citep{muno20}. The timescale that young NSs appear as FRB sources is $\tau_{\rm FRB}\sim 100~\rm years$, the mean burst rate $\dot{N}\sim 10~\rm year^{-1}$ of each FRB source (FRB 180916.J0158+65 has a high burst rate of $25~\rm year^{-1}$, Ioka \& Zhang 2020). Therefore, the detection rate density of FRBs can be approximately estimated to be
\begin{equation}
R_{\rm FRB} =R_{\rm CCSN}f\tau_{\rm FRB}\dot{N}\sim 10^{7}~\rm Gpc^{-3}year^{-1}.
\end{equation}
The distance of FRB 180916.J0158+65 is about $D=0.15$ Gpc \citep{chim20}, it is difficult for a far distance to detect such a FRB source. Assuming a unform distribution of such a FRB source, the detection rate of FRBs $\dot{n}=R_{\rm FRB}4\pi D^{3}/3\sim 10^{3}~\rm day^{-1}$. Based on the calculated number of the primordial black hole-NS collisions in a single galaxy \citep{abra09}, \citet{abra18} derived the occurrence rate of FRBs in the sky to be $10^{3}~\rm day^{-1}$. Our estimation is consistent with their result, and agrees the inferred value from observations \citep{cham16,cale17}. The radio emission originates from the open magnetic lines, and the last open field line is related to the light cylinder radius. We can simply estimate the duration of an FRB to be $\bigtriangleup t\sim r_{\rm lc}/c\sim 1~\rm ms$.

\section{Discussion and Summary}
In this Letter, we propose an alternative mechanism producing the periodic radio bursts detected from FRB 180916.J0158+65.
In our model, the debris disk possessing angular momentum larger than the central NS can precess the emitter of FRBs by the Lense-Thirring effect, producing a periodicity in observation. When the viscous parameter of the disk $\alpha=10^{-8}$, the spin period of the pulsar $P=1-20~\rm ms$,  the mass inflow rate of the disk $\dot{M}=0.5-1.5\times 10^{18}~\rm g\,s^{-1}$, the Lense-Thirring effect could result in a precession period $P_{\rm LT}=16.35\pm0.18~\rm days$. Recently, FRB 121102 was reported
to have a tentative bursting period of $159^{+3}_{-8}$ days \citep{rajw20}. If the observation is confirmed, our precession model can also account for such a bursting period. A fallback disk surrounding a NS that is more young than FRB 180916.J0158+65 possesses a high mass inflow rate, resulting in a long precession period (see also equation 8).

In our precession model, a new born NS would successively experience accretion and propeller phase because the mass inflow rate in the disk $\dot{M}\propto t_{\rm disk}^{-1.35}$. In the accretion and propeller phase, particle acceleration processes in the magnetospheric gap of the NS are quenched, the radio radiation turns off \citep{li06}. With the decrease of the mass inflow rate, the inner radius of the disk locates the outside the light cylinder, and the radio emission turns on. The young NS with millisecond period emits strong radio emission via a particle acceleration mechanism, and the radio luminosity depends on the magnetic field and spin period. The bursts are likely to be emitted from a fixed region on the NS such as a magnetic pole, which can produce a small viewing angle or large polar zone to ensure 4-day active window \citep{chim20}. However, the precession of a fallback disk causes a periodicity of FRBs. In view of the fact that the precession period is related to the spin period and the mass inflow rate (which depends on the disk mass and the age), most FRBs are difficult to detect repeating bursts due to a very long precession period.

The biggest problem of the Lense-Thirring precession model is the requirement of an extremely low viscous parameter $\alpha=10^{-8}$, which is 5-6 orders of magnitude lower than the canonical value $\alpha=0.01-0.001$. First, such a low viscosity is much higher than the molecular viscosity \citep{sara80}. Second, the MHD simulation also confirmed the existence of an extremely low viscous parameter $\alpha<10^{-5}$ \citep{hawl95}. Third, \citet{shak73} proposed a wide range as $10^{-15}(\dot{M}_{18}/2)^{2}<\alpha<1$, and our viscous parameter does still not exceed this scope. Therefore, our precession model remains marginal reliability.

Based on the precession model, we also constrain the fallback disk mass, the magnetic field of the NS, and bursting energy. A massive disk with a mass $\sim 10^{-3}~\rm M_{\odot}$ could be responsible for the detected periodicity in FRB 180916.J0158+65.
The emitter should be a young NS with an age of $\sim 10^{3}~\rm years$ and a normal magnetic field $10^{12}~\rm G$, in which the rotational energy of the NS can be visible as FRBs for $400~\rm years$. Furthermore, our model can also approximately interpret the occurrence rate of $\sim 10^{3}~\rm day^{-1}$ and the duration of $\sim 1$ ms of FRBs.

The idea that this source is a young NS seems to be in contradiction with the work performed by \citet{lyut17}. The local plasma would contribute to almost half of the DM for repeating FRB. Because the young NS should be enclosed by an expanding
supernova (SN) shell, the DM would sharply decrease with time, i. e. $DM\propto t^{-2}$ \citep{lyut16,piro16}. Therefore, nearly constant detection for the DM over few years can rule out the probability that young NSs become the FRB sources \citep{chat17}. First, the age of the NS is about $10^{5}~\rm years$ in our model, the change of the DM in a short duration can be ignored. Second, the NS may be far away from its birth place due to an extremely high kick velocity \citep{chat04,hobb05}. Some rotating young pulsars have not any surrounding SN ejecta, and non-relativistic pulsar wind nebula can produce synchrotron
radio emission to interpret persistent radio source associated with FRB 121102 \citep{dai17}. Certainly, our precession model requires a tilted fallback disk, and its misalignment probably originate from an asymmetrical supernovae explosion.

\begin{ack}
We are grateful to the referee Prof. Marek A. Abramowicz for helpful comments. We also thank Yun-Wei Yu, and Xiang-Dong Li for useful discussions on FRBs and fallback disks. This work was partly supported by the National Natural Science Foundation
of China (under grant number 11573016, 11733009), Program for Innovative Research Team (in Science and Technology)
in University of Henan Province, and China Scholarship Council.
\end{ack}


\end{document}